\documentclass[10pt,twocolumn,showpacs,preprintnumbers,amsmath,amssymb,aps,prb,
superscriptaddress,floatfix,footinbib]{revtex4-1}

\usepackage{graphicx}
\usepackage{dcolumn}
\usepackage{bm}

\begin{document}

\preprint{APS/123-QED}

\title{Exchange coupling constants at finite temperature}

 \author{S.\ Mankovsky}
 \affiliation{%
   Department  Chemie,  Physikalische  Chemie,  Universit\"at  M\"unchen,
   Butenandstr.\  5-13, 81377 M\"unchen, Germany\\}
\author{S.\ Polesya}
 \affiliation{%
   Department  Chemie,  Physikalische  Chemie,  Universit\"at  M\"unchen,
   Butenandstr.\ 5-13, 81377  M\"unchen, Germany\\}
\author{H.\ Ebert}
 \affiliation{%
   Department  Chemie,  Physikalische  Chemie,  Universit\"at  M\"unchen,
   Butenandstr.\  5-13, 81377 M\"unchen, Germany\\}

\date{\today}

\begin{abstract}
  An approach to account for the effect of thermal lattice vibrations
  when calculating exchange coupling parameters is presented on the
  basis of the KKR (Korringa-Kohn-Rostoker) Green function method making
  use of the alloy analogy model. Using several representative systems,
  it is shown that depending on the material the effect of thermal
  lattice vibrations can have a significant impact on the isotropic
  exchange as well as anisotropic Dzyaloshinskii-Moriya 
interactions (DMI). This should lead in turn to an additional
contribution to the temperature dependence of the magnetic properties of
solids, which cannot be neglected in the general case. As an example, we
discuss such an influence on the critical temperature of various
magnetic phase transitions. In particular, in the case of skyrmion
hosting materials, a strong impact 
of lattice vibrations on the DMI is an additional source for temperature
dependent skyrmion stability which should be taken into consideration. 
\end{abstract}

\maketitle

\subsection{INTRODUCTION}

The impact of finite
temperatures on the various physical material properties
is one of the most important issues in solid state physics
that is discussed in the literature with respect to
various aspects.
This holds in particular for finite temperature magnetic
and transport properties of materials calculated on an ab-initio level.
For that purpose, a very efficient approach -- the so-called
alloy analogy model -- has been introduced
recently \cite{EMC+15}, that allows to account for the impact
 of temperature induced lattice vibrations and spin fluctuations
 on linear response properties, as for example the electrical and
 spin conductivity, the Gilbert damping and others.
In these cases, the corresponding
response tensor  $\chi_{AB}$ may be written as
$\chi_{AB} \propto \, {\rm{Tr}}\, \big\langle A \, \Im G^+ \, B \, \Im G^+
\big\rangle_{\rm T}$,
where the operators $A$ and $B$ represent the
relevant observable and perturbation, respectively, while
$ G^+$ stands for the  retarded Green function  \cite{But85}.
Within the  alloy analogy model lattice vibrations  and spin fluctuations are
treated as uncorrelated, quasi-static atomic displacements and spin tiltings,
respectively,  with an amplitude depending on temperature.
Following the scheme used to calculate the residual
resistivity of disordered
alloys  \cite{Vel69, But85} by means of
the single-site Coherent Potential Approximation
(CPA), the thermal average $\langle ... \rangle_{ T}$
of a linear response quantity is obtained
as the configurational average
 over a set of appropriately  chosen set of  atomic
 displacements and spin tiltings using the CPA alloy theory
 \cite{EMKK11,MKWE13,EMC+15}.

The central idea
of the alloy analogy model was used
already previously to account for thermal magnetic disorder
when dealing with finite-temperature magnetic
properties  by means of  first-principles calculations 
done on the basis of the disordered local moment (DLM) model
 \cite{GPS+85,GBS+91,SG92b}.
This approach was formulated at the beginning
on a non-relativistic level. 
Its extension to the relativistic
disorder local moment (RDLM) model 
allowed  in particular 
to investigate the impact of thermal
spin disorder on the 
magneto-crystalline anisotropy (MCA)  \cite{SSB+06,BWS+07}.

So  far, most calculations of the exchange 
parameters have been
performed for ideal crystal structures 
assuming the {\em lattice temperature}
   $T_{\rm lat} = 0$~K.
Even for this situation, 
already a pronounced dependency
of the results  on the specific atomic positions 
could be observed for some cases \cite{BEH12,SU98}.
The significant influence of lattice vibrations on the magnon
excitations of fcc Fe has been reported for example by Sabiryanov and
Jaswal\cite{SJ99}, who calculated the exchange coupling parameters
accounting for corrections due to atomic displacements using a
frozen-phonon scheme. 
A substantial change for the exchange coupling
   parameters in bcc Fe was also reported 
   to be induced by a Burgers type lattice
   distortion which can be connected to the single $N$ point $TA_1$
   phonon mode \cite{MPE+13}.
   Recently, a strong impact of lattice vibrations on the electronic
   structure and magnetic properties of materials was shown employing the disordered
   local moments molecular dynamics (DLM-MD) method
   \cite{AKG+16,MABA18}. 
   This approach was  also used to investigate
   corresponding temperature induced changes of the exchange
   coupling  parameters,
   associated with thermal lattice vibrations \cite{RP18}.
Note that these
   DLM-MD calculations make use of supercell technique to simulate
   thermal atomic displacements in the system.   
     Di Gennaro et al. \cite{GMO+18} have investigated the combined
     effects of 'phononic' and 'magnonic' temperatures on the
spin-wave dispersion, stiffness, and Curie temperatures of Fe, Ni, and
permalloy by combining first-principles methods with model
Hamiltonians. Following the idea reported in Ref.\ \onlinecite{SJ99},
the authors take into account corresponding corrections to the
exchange parameters, associated with the thermal root-mean-square atomic
displacements at a given temperature.

Below we present a scheme to account  within the framework of  
the alloy analogy model 
for thermal lattice vibrations
when calculating exchange coupling  parameters.
As will be demonstrated by various examples,
such calculations can be done on the basis of
a ferromagnetic  state or 
a more realistic paramagnetic  DLM reference 
state.

\section{Theoretical background \label{TB}}

In the following the temperature dependence of the
parameters of the extended  Heisenberg Hamiltonian
\begin{eqnarray}
 H_{ex} &=&
 - \sum_{ij}
J_{ij} (\hat{e}_i \cdot \hat{e}_j)
 -  \sum_{ij} \vec{D}_{ij}[\hat{e}_i\times \hat{e}_j] \; .
\end{eqnarray}
will be considered.
Here $J_{ij}$ is the isotropic exchange coupling parameter
connected with the spin moments on sites $i$ and $j$ pointing
along the directions $\hat{e}_i$ and $\hat{e}_j$, respectively,
 while
$\vec{D}_{ij}$ represents the Dzyaloshinskii-Moriya (DM) interaction.
We will focus first of all on the properties of the isotropic
exchange parameters $J_{ij}$,
which are given by the average over the  diagonal elements of the
exchange coupling
tensor \cite{USPW03}.
Making use  of relativistic multiple-scattering formalism
the elements of this tensor can be written
for
$T = 0 $~K as \cite{EM09a}
\begin{eqnarray}
 J_{ij}^{\alpha_i \alpha_j}  &=&  -\frac{1}{2\pi} \Im \int dE\, \mathrm{Trace}\,
\Delta \underline{V}^{\alpha_i} \underline{\tau}^{ij}
\Delta \underline{V}^{\alpha_j} \underline{\tau}^{ji} \; .
\label{eq:Jij_stand}
\end{eqnarray}
Here $\underline{\tau}^{ij}$ is the so-called scattering path operator
connecting  sites $i$ and $j$
with the underline indicating matrices in the
$\Lambda=(\kappa,\mu)$-representation\cite{Ros61}. The
corresponding on-site coupling for site $i$
is represented by the matrix
%
\begin{eqnarray}
\Delta V^{i\alpha}_{\Lambda\Lambda'} & = & \int d^3r
Z^{\times}_{\Lambda}(\vec{r})\beta \sigma_{\alpha}B(r)
Z_{\Lambda'}(\vec{r}) \; ,
\label{eq:Jij_ME}
\end{eqnarray}
where $\alpha$ is one of the standard
Dirac matrices, $\sigma_{\alpha}$ is a $4 \times 4$-Pauli matrix\cite{Ros61}
and $B(r)$ is the spin-dependent part of the exchange-correlation
potential set up within local spin-density theory\cite{EM09a}.
Finally, the wave functions $Z_{\Lambda}(\vec{r})$ are solutions
to the Dirac equation normalized according to the
relativistic multiple-scattering formalism \cite{EBKM16}.

To apply the  expression in Eq.\ (\ref{eq:Jij_stand})
for the case of lattice vibrations at finite temperatures,
 we use again the alloy analogy model based on
 the adiabatic approximation.
 This implies that a discrete set of $N_{v}$ displacement
vectors $\Delta \vec{R}^q_v(T)$ with probability $x^q_v$
($v=1,..,N_{v}$) is constructed for each basis atom $q$
within the crystallographic unit cell.
The vectors  $\Delta \vec{R}^q_v(T)$
 are connected with the temperature
dependent root mean square displacement $(\langle u^2\rangle_T)^{1/2}$
according to the relation:
%
\begin{equation}
\label{eq:displacement}
\sum_{v=1}^{N_{v}} {x^q_{v}} | \Delta \vec{R}^q_v(T) |^2 = \langle u_q^2\rangle_T \;.
\end{equation}
%
For the applications presented below,
the temperature dependent
 root mean square displacement is estimated using  Debye's theory,
providing a simple connection between  $\langle
u_q^2\rangle_T$ and the {\em lattice temperature}.

Each displacement vector $\Delta \vec{R}_v(T)$
determines a corresponding   U-matrix
 $\underline{U}_v$ that describes
for all matrices in the  $\Lambda$-representation
 the coordinate transformation
 from a shifted atom position to the original
 equilibrium position.
This allows in particular to connect the single-site
t-matrix   $\underline{t}_v$ for a shifted atom
to the common global frame of reference used by the
multiple scattering calculations.
Within the alloy analogy model, each member in the set
of  $N_{v}$ displacement vectors $\Delta \vec{R}_v(T)$
can now be  seen as a pseudo-component of a multi-component  pseudo alloy.
As for a substitutional alloy,
the site diagonal configurational
average can this way be determined by solving the
multi-component CPA equations
referring to the global frame of  reference:
%
\begin{eqnarray}
\label{eq:CPA1}
\underline{\tau}_{{\rm CPA}} &= &
\sum_{v=1}^{N_{v}}
x_v \underline{\tau}_{v}
\\
%
\label{eq:CPA2}
\underline{\tau}_{v}& = &
\big[
    (\underline{t}_{v})^{-1}
-   (\underline{t}_{{\rm CPA}})^{-1}
+   (\underline{\tau}_{{\rm CPA}})^{-1}
\big]^{-1}
\\
\label{eq:CPA3}
\underline{\tau}_{{\rm CPA}}
 & = & \frac{1}{\Omega_{{\rm BZ}}}
 \int_{\Omega_{\rm BZ} } d^{3}k
\left[ (\underline{t}_{{\rm CPA}})^{-1}
      - \underline{G}(\vec{k},E)  \right]^{-1}  \; ,
\end{eqnarray}
%
where the CPA medium is described by a corresponding
CPA single-site t-matrix $\underline{t}_{{\rm CPA}}$
and scattering path operator $\underline{\tau}_{{\rm CPA}}$.
The first of these equations expresses  the requirement for the mean-field CPA
medium that embedding of a component $v$ into the medium
should not lead in the average to an additional scattering,
with Eq.\ (\ref{eq:CPA2}) giving the corresponding  scattering path operator
$\underline{\tau}_{v}$ for the
embedded component $v$.
Finally,  Eq.\ (\ref{eq:CPA3}) gives   $\underline{\tau}_{{\rm CPA}}$
 by a Brillouin zone
integral in terms of   $\underline{t}_{{\rm CPA}}$ and
 the so-called KKR structure constants $\underline{G}(\vec{k},E) $ \cite{EKM11}.

 Assuming -- in line with the adiabatic approximation -- a frozen potential for the
displaced atoms and neglecting
correlations between the atomic displacements,
Eqs.\ (\ref{eq:CPA1}) to (\ref{eq:CPA3})
allow to  evaluate  of the necessary  thermal
configurational averaging when dealing with  Eq.\ (\ref{eq:Jij_stand})
for finite temperatures. This way one gets
for the temperature dependent exchange coupling constants:
\begin{eqnarray}
 \bar{J}_{ij}^{\alpha_i \alpha_j}  &=&  -\frac{1}{2\pi} \Im  \int dE \, \mathrm{Trace}
\langle \Delta \underline{V}^{\alpha_i} \underline{\tau}^{ij} \Delta \underline{V}^{\alpha_j} \underline{\tau}^{ji} \rangle_c \; ,
\label{eq:Jij-T_1}
\end{eqnarray}
where $\langle ... \rangle_{\rm c}$
represents  the configurational average with respect to the set of displacements.
In all calculations we have used a set of $N_{v} = 14$ displacements as
increasing $N_{v} $ led only to  minor changes to the final results.
As discussed for example in the context of the electrical conductivity,
dealing with a configurational average as occurring in Eq.\ (\ref{eq:Jij-T_1})
leads to the so-called vertex corrections\cite{Vel69,But85}.
As the expression in Eq.\ (\ref{eq:Jij-T_1}) refers explicitly to a
specific pair of sites, these have been ignored here;
i.e.\ the
configuration average  has been simplified to
$\langle {\Delta \underline V}^{\alpha} \, \ \underline{\tau}^{ij} \, \rangle_{\rm c}  \langle
         {\Delta \underline V}^{\beta} \, \ \underline{\tau}^{ji}  \rangle_{\rm c} $.

\section{Computational details}
\label{SEC:Computational-scheme}

The results presented below are based on self-consistent
first-principles electronic structure
calculations performed using the spin-polarized
relativistic Korringa Kohn Rostoker  Green function (SPR-KKR-GF) method
\cite{SPR-KKR7.7,EKM11}, using the atomic sphere approximation (ASA).
The local spin density approximation (LSDA) to spin density
functional theory (SDFT) has been used with a parametrization for the
exchange and correlation potential as given by Vosko {\em et  al.}
\cite{VWN80}. For the angular momentum expansion of the Green function
the angular momentum cutoff $l_{\rm max} = 3$ was used.
Within the present work, the following systems have been considered:
 bcc Fe ($a = 5.40$~a.u.), fcc Ni ($a = 6.65$~a.u.),
 ferromagnetic ($a = 5.66$~a.u.) and  anti-ferromagnetic ($a = 5.63$~a.u.)  B2 FeRh,
  1ML Fe on the (111) surface of Pt ($a = 7.40$~a.u. for fcc Pt), and 1ML
  Fe on the (111) surface of Au ($a = 7.68$~a.u. for fcc Au) 
with the corresponding structure parameters given in atomic units, i.e.\
as multiples of the Bohr radius, in parentheses.
The calculations
for 1ML Fe/Pt(111) and 1ML Fe/Au(111)  have been performed using a
supercell geometry with a (1ML Fe/3ML Pt(Au)/5ML ES) 
supercell (where ES stands for empty sphere), with
Fe occupying ideal fcc positions, i.e.\ without optimization of the
interlayer distance. A k-mesh with $25 \times 25 \times 25$ grid points
was used for the integration over the BZ of the three-dimensional 
bulk systems and with $46 \times 46 \times 5$ grid points for
1ML Fe on the (111) surface of Pt or Au, respectively. For
the calculations of the exchange parameters as a function of the
occupation the corresponding energy integration has been performed using
an energy mesh with 200 energies having a constant imaginary part
of 1 meV.

\section{Results \label{res}}

As it is mentioned above, one may expect
that the  modification of the electronic structure
due to  thermal lattice vibrations
will not only influence transport and other response
properties, but also  the exchange coupling parameters.
That this is indeed the case is demonstrated
in the following for
the  elemental ferromagnets bcc Fe and fcc Ni,
B2 bulk FeRh, as well as for a Fe monolayer on
Pt(111)
as representative examples.
\medskip

\begin{figure}[htb]
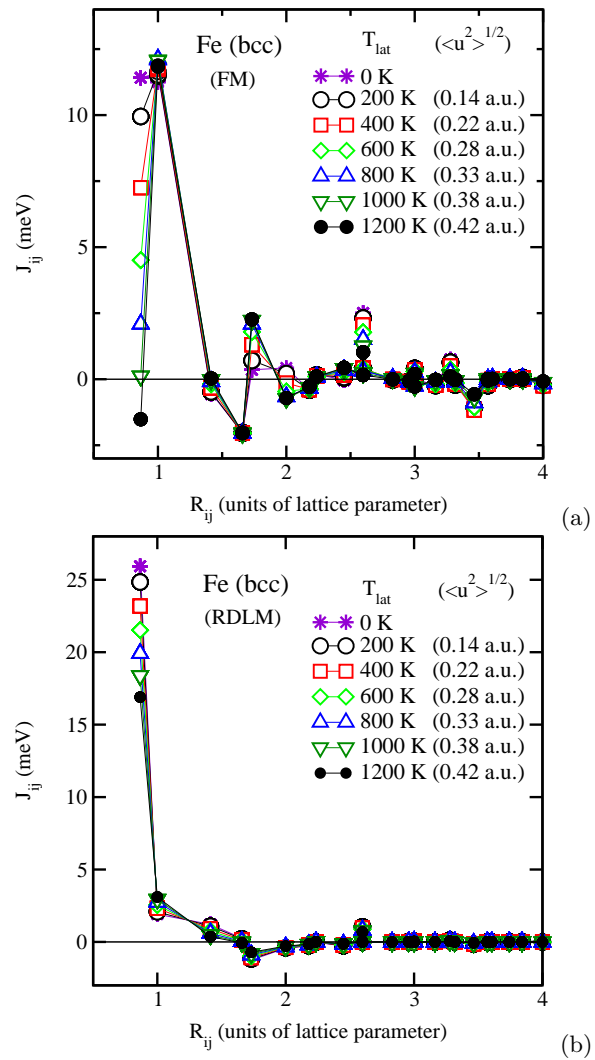

\includegraphics[width=0.4\textwidth,angle=0,clip]{CMP_FM_Fe_JXC_vs_T_vib_mod.eps}\;(a)
\includegraphics[width=0.4\textwidth,angle=0,clip]{CMP_RDLM_Fe_JXC_vs_T_vib_mod.eps}\;(b)
\caption{\label{fig:Fe_JXC_FM-DLM_vibra} The isotropic exchange coupling
  parameters $J_{ij}$ for bcc Fe calculated for the FM (a) and  DLM (b)
  reference states. The results are represented for different amplitudes
  of the thermal lattice vibrations given in terms of
 the   rms displacement
$(\langle u^2\rangle_T)^{1/2}$ and corresponding
{\em lattice temperature} $T_{\rm lat}$.
    }
\end{figure}

\begin{figure}[htb]
\includegraphics[width=0.4\textwidth,angle=0,clip]{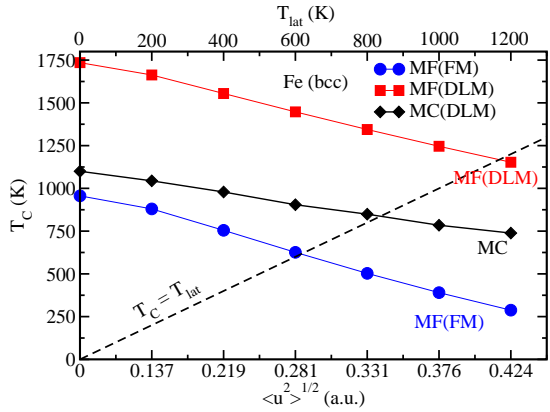}\;
\caption{\label{fig:Fe_TC_vs_vibra} Theoretical Curie temperature
$T_{\rm C}$
  plotted as a function of the amplitudes of thermal lattice vibrations
$(\langle u^2\rangle_T)^{1/2}$
   calculated for the FM  and
  DLM reference states either using the MFA or MC simulations
together with the relation
between the
{\em lattice temperature} $T_{\rm lat}$
 and
$(\langle u^2\rangle_T)^{1/2}$.
    }
\end{figure}

The isotropic exchange coupling parameters
$J_{ij}$ calculated for the FM
reference state of bcc Fe
are plotted in Fig.\ \ref{fig:Fe_JXC_FM-DLM_vibra} (a)
for different amplitudes of thermal lattice vibrations
related to
 a corresponding {\em lattice temperature} $T_{\rm lat}$
 according to the Debye
model.
%
%
As one can see, there are indeed
pronounced modifications of the exchange coupling
parameters due to the lattice vibrations
that depend strongly on the considered pair of sites.
By far the most significant changes  are found for
 the nearest-neighbor interaction parameters
 that
  decrease strongly with an  increase
of the amplitude of the thermal displacements
or the {\em lattice temperature}, respectively.
This in turn should
have a corresponding impact on the
Curie temperature  $T_{\rm C}$.
Within the   mean-field  approximation (MFA),  $T_{\rm C}$
is essentially given by a sum
$\sum_j J_{ij}$ over the coupling parameters
allowing therefore in a simple way  to monitor
the dependency of   $T_{\rm C}$ on the
effectice  {\em lattice temperature} $T_{\rm lat}$
or, equivalently, on the
temperature dependent rms displacement
$(\langle u^2\rangle_T)^{1/2}$.
Fig.\ \ref{fig:Fe_TC_vs_vibra} (circles) shows corresponding
results for  $T_{\rm C}$ as a function of
$(\langle u^2\rangle_T)^{1/2}$
obtained by summing $J_{ij}$ within a
sphere with radius $R_{\rm max} = 5a$, with $a$ being the lattice
parameter.

 Fig.\ \ref{fig:Fe_TC_vs_vibra} (circles) shows corresponding
results for $T_{\rm C}$
obtained via the MFA  as a function of
the temperature dependent
 rms displacements
$(\langle u^2\rangle_T)^{1/2}$.
%
%
Keeping in mind that the mean field approximation (MFA) normally overestimates the
critical temperature when compared to
results obtained from Monte Carlo (MC)
simulations  or RPA
(random phase approximation)
 based calculations,
one  notes that the MFA result for
$T_{\rm C}$ of bcc Fe, evaluated without  accounting for the
 lattice vibrations,
is rather close to the  experimental value, $T^{\rm exp}_{\rm C} = 1043$~K. However,  a finite amplitude of the
lattice vibrations leads to a significant monotoneous 
decrease of  $T^{\rm MF}_{\rm C}$ with $(\langle u^2\rangle_T)^{1/2}$ 
implying a corresponding deviation from  experiment.
As  mentioned above, more reliable results for the Curie
temperature can be obtained
on the basis of the exchange coupling parameters
calculated for the PM reference state described here within the disordered
local moment (DLM) approximation.
 Using the non-relativistic version of this model, magnetic
disorder  in the PM state is accounted for
by averaging over all possible directions of the spin moments.
Equivalent to this,  is to
consider a
pseudo alloy  $Fe_{0.5}^{\rm up}Fe_{0.5}^{\rm down}$
of Fe atoms
with  opposite spin
moments oriented up and down, respectively.
Fig.\ \ref{fig:Fe_JXC_FM-DLM_vibra} (b) gives the corresponding  exchange
coupling parameters of Fe
for the PM reference state on the basis  of the DLM Model.
These parameters and their temperature dependence
are quite different
from those obtained for the FM reference state. As a consequence,
the corresponding MFA Curie temperature ($\approx 1700$~K )  exceeds
the value obtained for the FM reference state in an appreciable way
when thermal lattice vibrations are ignored.
This observation was already reported  in the literature before (see e.g.\ \onlinecite{BSW08}).
A finite amplitude of the  thermal atomic displacements
leads again to a lower  MFA-based Curie temperature, as it is
shown in Fig.\ \ref{fig:Fe_TC_vs_vibra}, 
reaching the value $T_{\rm C}^{\rm MF} \approx 1200$~K 
when requiring that the Curie temperature  and {\em lattice temperature}
coincide.

Fig.\ \ref{fig:Fe_TC_vs_vibra} (triangles) gives also 
results for the Curie temperature 
obtained by  MC simulations considering 15 atomic shells
around each atom 
using DLM-based exchange parameters. 
In this case, the Curie temperature $T_{\rm C}^{\rm MC}$, calculated
for an unperturbed lattice slightly overestimates the 
experimental value. When the amplitude of thermal lattice vibrations
increases,   $T_{\rm C}^{\rm MC}$ also goes down and coincides 
with the {\em lattice temperature} $T_{\rm lat}$ at around $1000$~K
underestimating slightly  the  experimental Curie temperature this way.
This small deviation might among others 
be ascribed to the  approximate treatment 
of lattice vibrations when calculating $J_{ij}$ 
that in particular neglects correlations in the thermal motion of the atoms.
\medskip

\begin{figure}[htb]
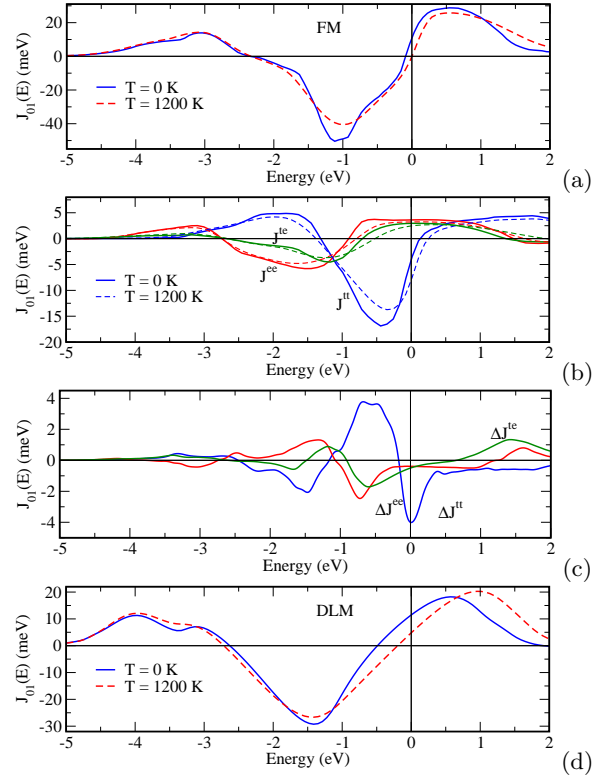

\includegraphics[width=0.4\textwidth,angle=0,clip]{Fe_J_01_vs_E_and_T_FM_mod.eps}\;(a)
\includegraphics[width=0.4\textwidth,angle=0,clip]{Fe_J_01_vs_E_and_T_FM_orbital.eps}\;(b)
\includegraphics[width=0.4\textwidth,angle=0,clip]{Fe_J_01_vs_E_and_T_FM_Delta_orbital.eps}\;(c)
\includegraphics[width=0.4\textwidth,angle=0,clip]{Fe_J_01_vs_E_and_T_DLM_mod.eps}\;(d)
\caption{\label{fig:Fe_JXC_vs_E_T_vibra_FM} The occupation dependendence
of the exchange
  coupling parameter $J_{01}$ of bcc Fe  for the FM (a) and  the DLM (c)
  reference states. Dashed line represents  results for
the  {\em lattice temperature} $T_{\rm lat}= 1200$~K.
  (b) represents the orbital-resolved parameters for
  the FM  reference state, ${\tilde J}^{tt}_{01}$, ${\tilde J}^{ee}_{01}$ and
  ${\tilde J}^{te}_{01}$, respectively, while (c) gives their changes due to thermal lattice vibrations when increasing $T_{\rm lat}$ from 0 to $1200$~K.
    }
\end{figure}

To get more insight concerning the temperature
dependence of the exchange coupling parameters, Fig.\
\ref{fig:Fe_JXC_vs_E_T_vibra_FM} (a) shows the nearest neighbor
parameter $J_{01}$ for FM bcc Fe for two different temperatures
as a function of the upper limit of the energy integration 
 in Eq. (\ref{eq:Jij_stand}) (with $E = 0$ eV the true Fermi energy)
reflecting its dependence on the occupation.
%
The solid and dashed lines represent results obtained 
without and with lattice vibrations, respectively, accounted for. One
can see, that depending on the occupation of the valence band the lattice
vibrations can result either in a decrease or increase of the exchange
parameter.
Following  Kvashnin {\em et al.} \cite{KCS+16}, one can further 
decompose  $J_{ij}$ into its orbital
contributions. For the orbitals grouped according to
the representations of the cubic point group, $t_{2g}$ and $e_{g}$, the
exchange parameter can be decomposed 
according to the expression $J_{ij} = J^{t_{2g}-t_{2g}}_{ij}  +
J^{e_{g}-e_{g}}_{ij} + J^{t_{2g}-e_{g}}_{ij}$
allowing to monitor the dependence of
the  individual orbital contributions to $J_{ij}$ \cite{RP18}
on the lattice vibrations. 
In Fig.\ \ref{fig:Fe_JXC_vs_E_T_vibra_FM} (b) 
representative results are shown for  
the contributions of 
the  $l = 2, m = \pm 1$ ($t_{2g}$) and  $l = 2, m = 0$ orbitals ($e_{g}$)
to the nearest neighbor interaction parameter $J_{01}$,
with the corresponding representations given in parentheses.
To distinguish these data from those connected with  the complete
set of the cubic point group representations, $t_{2g}$ and $e_{g}$, we use the
symbol ${\tilde J}$ instead of $J$.
For calculations done without lattice vibrations ($T_{\rm lat}= 0$~K),
this decomposition reveals an antiferromagnetic
character for  the ${\tilde J}^{tt}_{01}$ parameter 
 in contrast to the ferromagnetic character of ${\tilde J}^{ee}_{01}$ and
${\tilde J}^{te}_{01}$. 
This finding is  in full agreement with previous work \cite{KCS+16,RP18}.
The change of the orbital resolved coupling parameters 
 ${\tilde J}^{\gamma \gamma'}_{01}$ ($\gamma (\gamma') =
 e \equiv e_{g},\,t \equiv t_{2g})$
when going from 0 to 1200~K is shown in Fig.\
\ref{fig:Fe_JXC_vs_E_T_vibra_FM} (c). 
Obviously, the most pronounced changes are found for the contribution 
 ${\tilde J}^{tt}_{01}$.
The observed changes are primarily ascribed  to the broadening of
 the electronic states due to the thermal lattice vibrations, 
leading either to an increase or decrease of ${\tilde J}^{\gamma \gamma'}_{01}$
or $J_{01}$, respectively,
 depending on the occupation of the energy band.
Finally,
Fig.\ \ref{fig:Fe_JXC_vs_E_T_vibra_FM} (d) represents results
obtained for the DLM reference state.
The electronic states in this case are broadened in addition due to the
thermally induced magnetic disorder in the system. 
Including thermal lattice vibrations in addition with  $T_{\rm lat}= 1200$~K
leads for the $J_{01}$ parameter   to changes w.r.t.\ 
$T_{\rm lat}= 0$~K comparable to those found for the 
ferromagnetic reference state (see Fig.\ \ref{fig:Fe_JXC_vs_E_T_vibra_FM} (a))

\medskip

\begin{figure}[htb]
\includegraphics[width=0.4\textwidth,angle=0,clip]{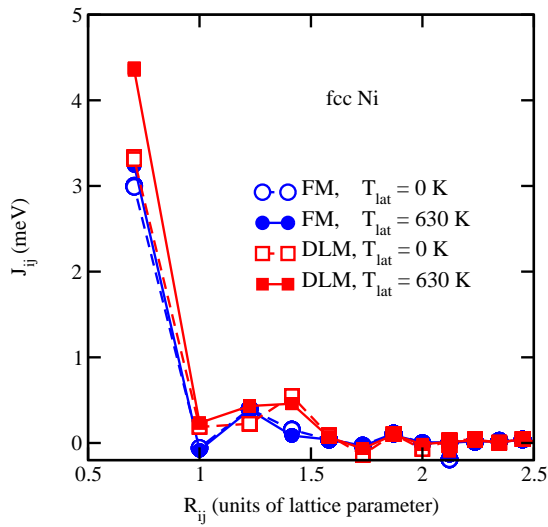}\;
\caption{\label{fig:Ni_JXC_FM-DLM_vibra}  Exchange coupling parameters
  $J_{ij}$ calculated for Ni for the FM and DLM reference state without
  lattice vibrations and  accounting for lattice vibrations corresponding to $T_{\rm lat} = 630$~K.
    }
\end{figure}

\begin{figure}[htb]
\includegraphics[width=0.4\textwidth,angle=0,clip]{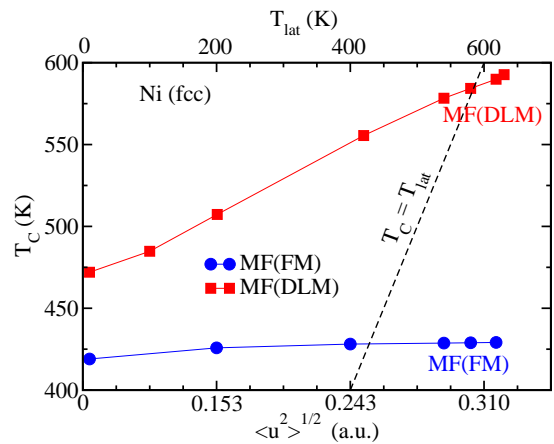}
\caption{\label{fig:Ni_JXC_vs_T}  The MF Curie temperature calculated
  for fcc Ni for the FM  and   DLM reference states, plotted as a
  function of the amplitudes of   thermal lattice vibrations  given in
  terms of {\em lattice temperature}.
    }
\end{figure}

The isotropic
  exchange coupling parameters $J_{ij}$ calculated for fcc
Ni  are shown in Fig.\
\ref{fig:Ni_JXC_FM-DLM_vibra}. 
%
%
For this material the
lattice vibrations lead to a tiny modification of the exchange
parameters calculated for the FM reference state as
the results 
for  $T_{\rm lat} = 0$ (circles)   and $630$~K  (triangles)
shown in  Fig.\ \ref{fig:Ni_JXC_FM-DLM_vibra} demonstrate.
The mean-field Curie temperature evaluated with these parameters
increases from $T_{\rm C}^{\rm MF} \approx 420$~K 
obtained with the  parameters for
 the unperturbed ground state ($T_{\rm lat} = 0$~K)
to  $T_{\rm C}^{\rm MF} \approx 430$~K 
for the state with an amplitude of lattice
vibrations corresponding to $T_{\rm lat} = 630$~K.
The well known itinerant-electron character of 
magnetism in Ni
leads  -- in  contrast to Fe --
for the  PM state above the Curie temperature
 to a very small or 
 vanishing magnetic moment (see e.g.\ Ref.\ \onlinecite{RKMJ07} 
and references therein). 
This prevents to perform standard self-consistent DLM
calculations as these also lead to a zero local magnetic moment 
for the paramagnetic DLM state.
For that reason, Ruban {\em et al.}
suggested to use a constrained local
exchange field when dealing with the magnetic properties of Ni.
As the subtle temperature dependent magnetism  of Ni is not the 
central issue of the present work,
we investigated  the simultaneous impact of lattice
vibrations and magnetic disorder on the $J_{ij}$ parameters
by performing the DLM-like calculations with the 
 spin moment constrained by using a frozen potential \cite{EMKK11,MKWE13,EMC+15}.
The resulting exchange coupling parameters calculated 
for the  DLM reference state
without account for lattice vibrations are given in Fig.\
\ref{fig:Ni_JXC_FM-DLM_vibra} by open squares, while
closed squares represent
 data for the  {\em lattice temperature} $T_{\rm lat} = 630$~K.
As one notes, the first-neighbor exchange parameters
significantly
increase with the temperature increase
as can be seen  in Fig.\ \ref{fig:Ni_JXC_FM-DLM_vibra}.
 The corresponding MFA Curie temperature shown in  Fig.\ \ref{fig:Ni_JXC_vs_T} by squares increases
from $\sim 470$~K for $T_{\rm lat} = 0$~K to $\sim 600$~K for
$T_{\rm lat} = 630$~K.
%
%
However, one should keep in mind that the MFA results lead 
usually  to an 
overestimation of the Curie temperature.
On the other hand, 
performing instead MC simulations  based on  the DLM derived
 exchange parameters calculated for
$T_{\rm lat} = 630$~K, leads to a
  Curie temperature $T_{\rm C} = 430$~K that is far below the experimental value.

\begin{figure}[htb]
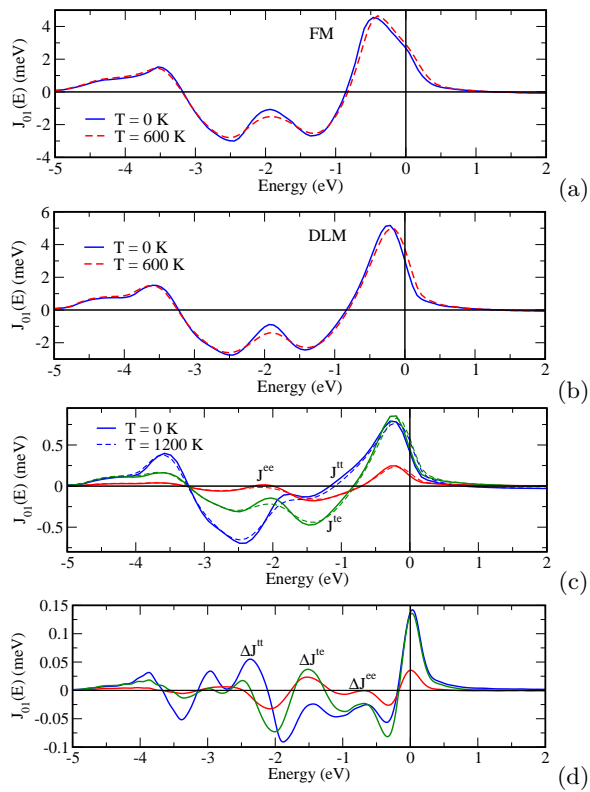

\includegraphics[width=0.4\textwidth,angle=0,clip]{Ni_J_01_vs_E_and_T_FM_mod.eps}\;(a)
\includegraphics[width=0.4\textwidth,angle=0,clip]{Ni_J_01_vs_E_and_T_DLM_mod.eps}\;(b)
\includegraphics[width=0.4\textwidth,angle=0,clip]{Ni_J_01_vs_E_and_T_DLM_orbital.eps}\;(c)
\includegraphics[width=0.4\textwidth,angle=0,clip]{Ni_J_01_vs_E_and_T_DLM_Delta_orbital.eps}\;(d)
\caption{\label{fig:Ni_JXC_vs_E_T_vibra_DLM} The occupation dependent exchange
  coupling parameter $J_{01}$ for fcc Ni for the FM (a) and  the DLM (b)
  reference states. (c) represents the orbital-resolved parameters for
  the DLM reference state, ${\tilde J}^{tt}_{01}$, ${\tilde J}^{ee}_{01}$ and
  ${\tilde J}^{te}_{01}$ and (d) their changes due to thermal lattice vibrations.
    }
\end{figure}

The occupation dependence of the  exchange coupling
parameter $J_{01}$ of Ni
 calculated for the FM reference state is shown  in
Fig.\ \ref{fig:Ni_JXC_vs_E_T_vibra_DLM} (a) for the two 
{\em lattice temperatures} 
$T_{\rm lat} = 0$ and $630$~K.
%
%
As to be expected from 
Fig.\ \ref{fig:Ni_JXC_FM-DLM_vibra}
 a relatively weak impact of thermal lattice vibrations
 is found in this case.
This can partially  be attributed to the rather low
critical temperature, i.e.\ temperature regime to be considered, 
for which the mean-square displacements of the atoms are
still too small to lead to significant changes in the electronic
structure. 
In line with this,
the temperature dependence 
of the parameter  for the DLM reference
state shown in Fig.\ \ref{fig:Ni_JXC_vs_E_T_vibra_DLM} (b) 
is found to be very similar to that for the FM state.
The orbital decomposition of the data 
for the DLM reference state 
that is given  in
Fig.\ \ref{fig:Ni_JXC_vs_E_T_vibra_DLM} (c) shows that all components
${\tilde J}^{tt}_{01}$, 
${\tilde J}^{ee}_{01}$ and 
${\tilde J}^{te}_{01}$ are positive for the occupation corresponding to
the true Fermi energy of fcc Ni and that for Ni
the most pronounced impact of lattice vibrations  occurs
 for the
${\tilde J}^{tt}_{01}$ and ${\tilde J}^{te}_{01}$ contributions.

\begin{figure}[htb]
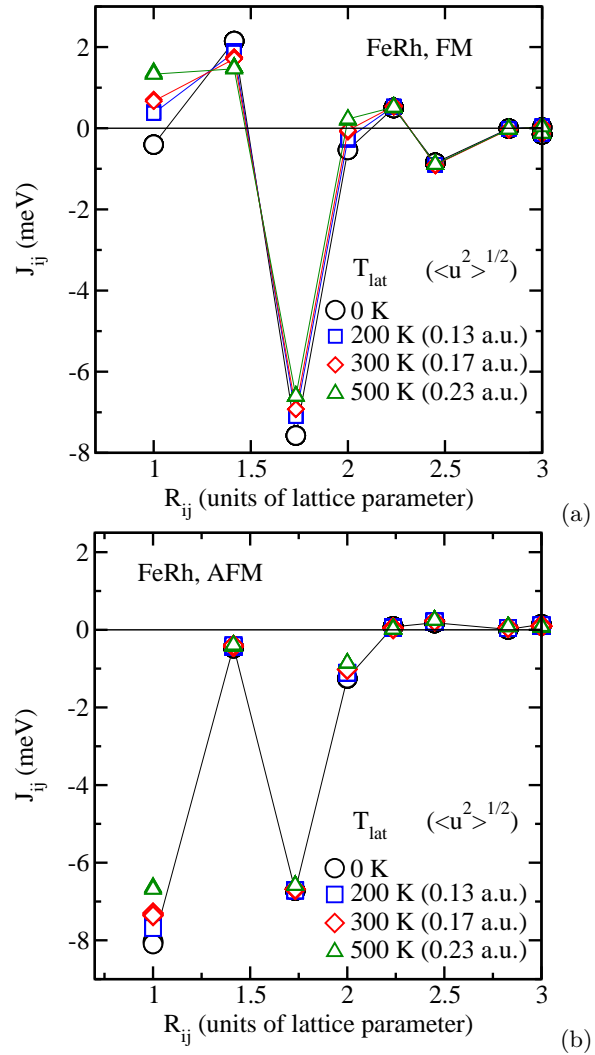

\includegraphics[width=0.4\textwidth,angle=0,clip]{CMP_Jij_vs_T-vibra_FeRh_FM_mod.eps}\;(a)
\includegraphics[width=0.4\textwidth,angle=0,clip]{CMP_Jij_vs_T-vibra_FeRh_AFM_mod.eps}\;(b)
\caption{\label{fig:FeRh_J-T} Interatomic Fe-Fe exchange coupling parameters
  corresponding to various temperatures, calculated for FeRh
  with the FM (a) and AFM (b) structures. 
  The temperature dependency is only due to the
   thermal lattice vibrations. The mean-square
  displacements corresponding to the considered
   temperatures are as follows:
  $0.13$~a.u.\ (200 K), $0.17$~a.u.\  (300 K), 
  and $0.23$~a.u.\   (500 K).
    }
\end{figure}

As an example for a compound,  the well known B2 
FeRh system  
that  exhibits a temperature induced AFM to FM
transition is considered in the following. 
According to first-principles calculations\cite{PMK+16}, the
metamagnetic transition can be seen as a result of 
the competition of Fe-Fe exchange interactions including 
indirect Fe-Rh-Fe interactions, 
which depend on the magnetic configuration. 
However, a
possible influence of lattice vibrations on the 
finite temperature magnetic properties of FeRh 
has not been discussed so far.
Within the present work, 
 calculations have been performed for the FM and AFM
configurations separately considering several values of 
{\em lattice temperatures}.
 The  corresponding results are given in
  Fig.\ \ref{fig:FeRh_J-T} for the FM (a) and AFM (b) states.
%
%
One can see in both cases that the increase of the amplitude of 
the thermal
lattice vibrations results in an increase of the interatomic FM
exchange and a decrease of the AFM exchange interactions.
This implies that thermal lattice vibrations
should  decrease the stability of the
low-temperature AFM phase upon  heating via
the induced changes of the exchange parameters for the FM as well as
AFM state. 
This should result in a decrease by about 40 K (using the {\it
    lattice temperature} $T_{lat} = 300$ K) of the critical temperature
  of the AFM-FM metamagnetic phase transition, that follows from the
  MC simulations. Note however, that these calculations do no account
  for the impact of lattice vibrations on the Fe-Rh exchange interactions. 

\begin{figure}
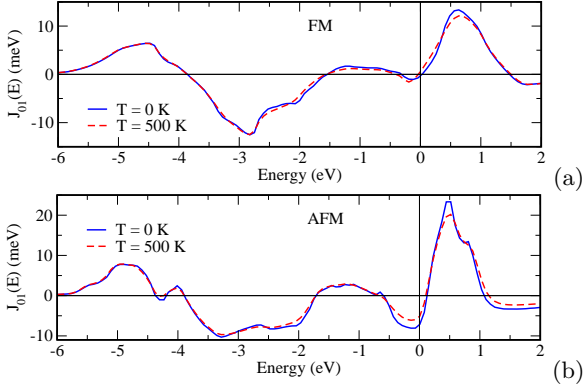

\includegraphics[width=0.4\textwidth,angle=0,clip]{FeRh_J_01_vs_E_and_T_FM.eps}\;(a)
\includegraphics[width=0.4\textwidth,angle=0,clip]{FeRh_J_01_vs_E_and_T_AFM.eps}\;(b)
\caption{\label{fig:FeRh_JXC_vs_E_T_vibra_FM} The occupation dependent Fe-Fe exchange
  coupling parameter $J_{01}$ for FeRh using the FM (a) and the AFM (b)
  structure as a reference states. The results are presented for two values of the
  {\em lattice temperature} $T_{\rm lat}$.
    }
\end{figure}

The occupation dependence of the 
 Fe-Fe exchange coupling parameter $J_{01}$ 
 of FeRh are shown in Fig.\
\ref{fig:FeRh_JXC_vs_E_T_vibra_FM}
for the FM as well as the  AFM reference states. 
%
%
One can see that the impact of lattice
vibrations on $J_{ij}$ is rather small 
over all occupation numbers or energies, repectively, 
 and is close to its maximum value
for the proper occupation number at the 
Fermi level, i.e.\ at $E=0$~eV.

\medskip

\begin{figure}
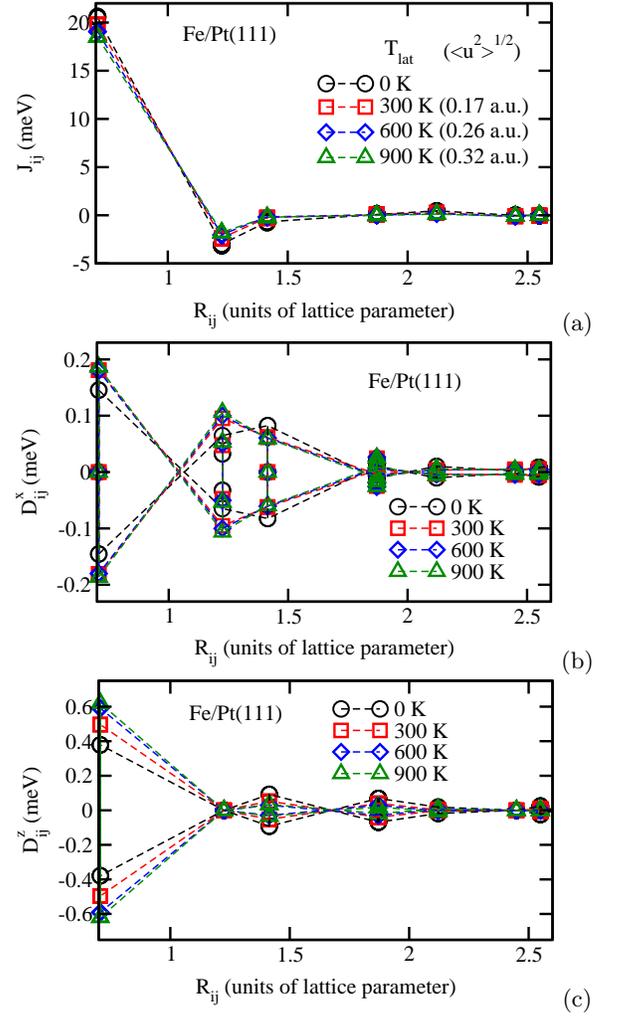

\includegraphics[width=0.4\textwidth,angle=0,clip]{CMP_Jij_Fe-Pt_mod.eps}\;(a)
\includegraphics[width=0.4\textwidth,angle=0,clip]{CMP_Dij_x_Fe-Pt_mod.eps}\;(b)
\includegraphics[width=0.4\textwidth,angle=0,clip]{CMP_Dij_z_Fe-Pt_mod.eps}\;(c)
\caption{\label{fig:FePt_J_DMIx} The isotropic
exchange
  coupling parameter $J_{ij}$ (a), the x-component $D^x_{ij}$ (b) and
  the z-component $D^z_{ij}$ (c) of the DMI for 1ML Fe on the Pt(111)
  surface for several values of the rms atomic displacement $(\langle
  u^2\rangle_T)^{1/2}$ (given in parentheses) corresponding to different {\em lattice
    temperatures} $T_{\rm lat}$.} 
\end{figure}

\begin{figure}
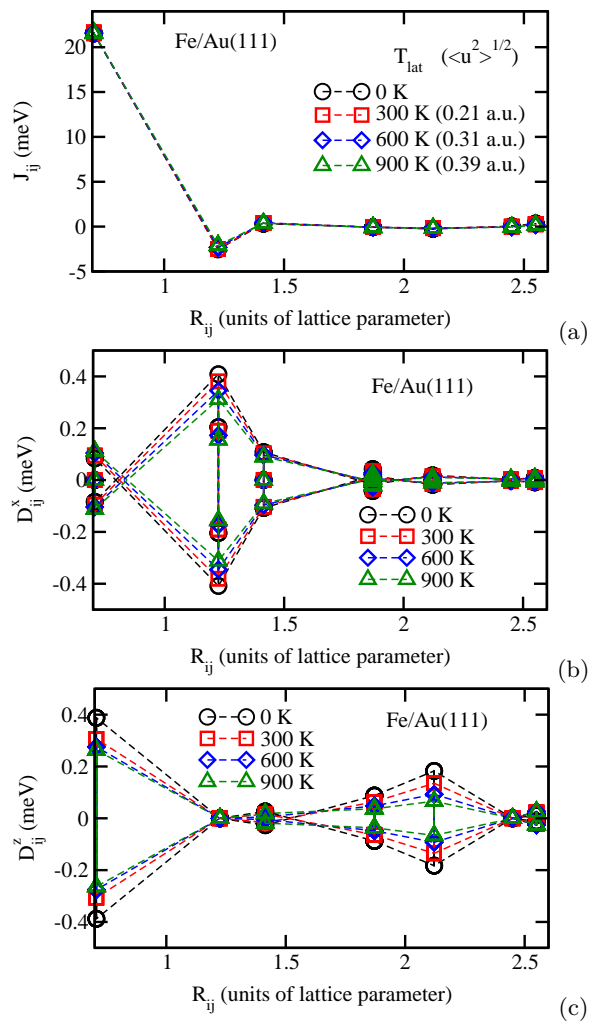

\includegraphics[width=0.4\textwidth,angle=0,clip]{CMP_Jij_Fe-Au_mod.eps}\;(a)
\includegraphics[width=0.4\textwidth,angle=0,clip]{CMP_Dij_x_Fe-Au_mod.eps}\;(b)
\includegraphics[width=0.4\textwidth,angle=0,clip]{CMP_Dij_z_Fe-Au_mod.eps}\;(c)
\caption{\label{fig:FeAu_J_DMIx} The isotropic
exchange
  coupling parameter $J_{ij}$ (a), the x-component $D^x_{ij}$ (b) and
  the z-component $D^z_{ij}$ (c) of the DMI for 1ML
  Fe/Au(111) for several values of the rms atomic displacement
  $(\langle u^2\rangle_T)^{1/2}$ (given in the parentheses) 
   corresponding
  to different {\em lattice temperatures} $T_{\rm lat}$.    }
\end{figure}

\begin{figure}
\includegraphics[width=0.3\textwidth,angle=270,clip]{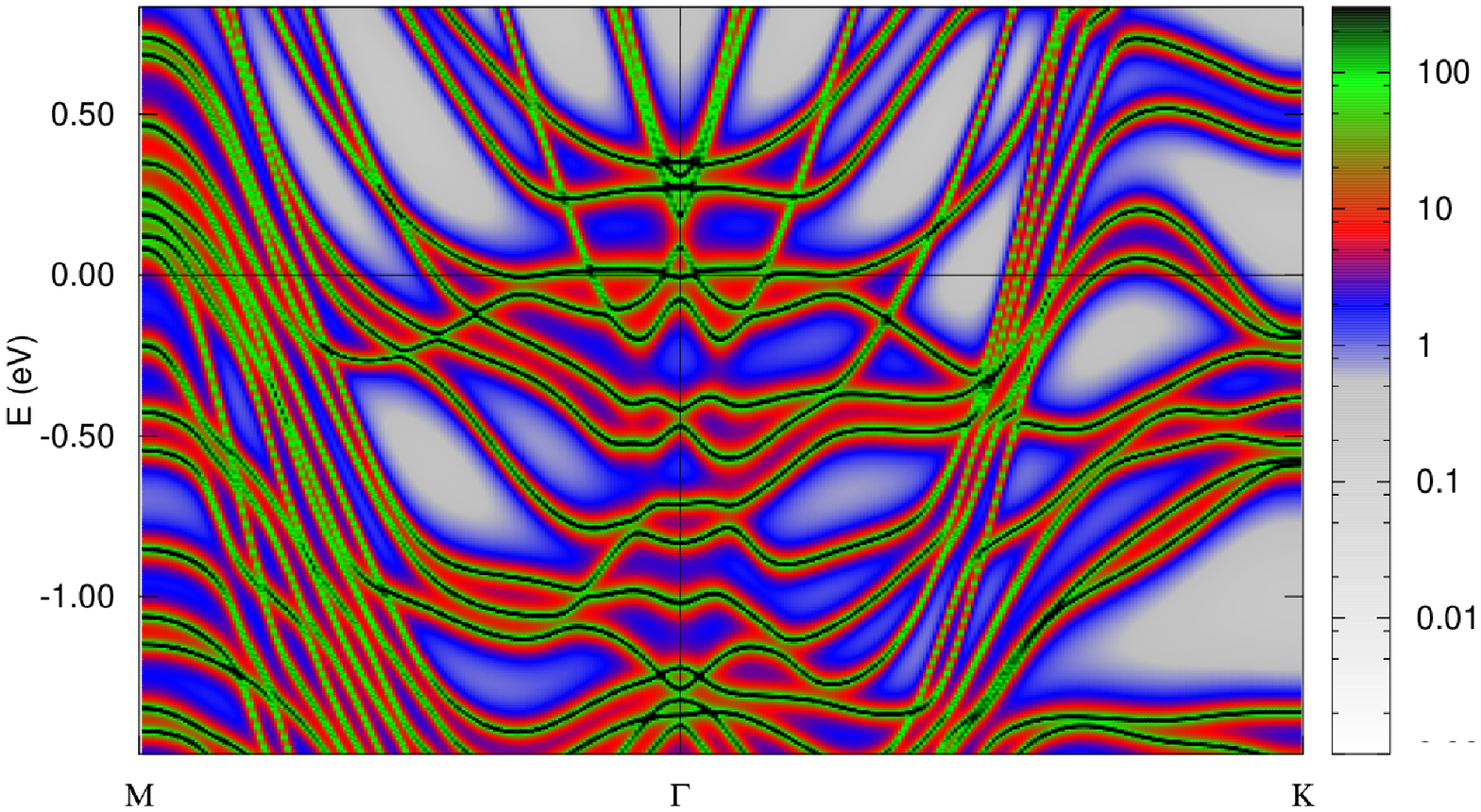}\;(a)
\includegraphics[width=0.3\textwidth,angle=270,clip]{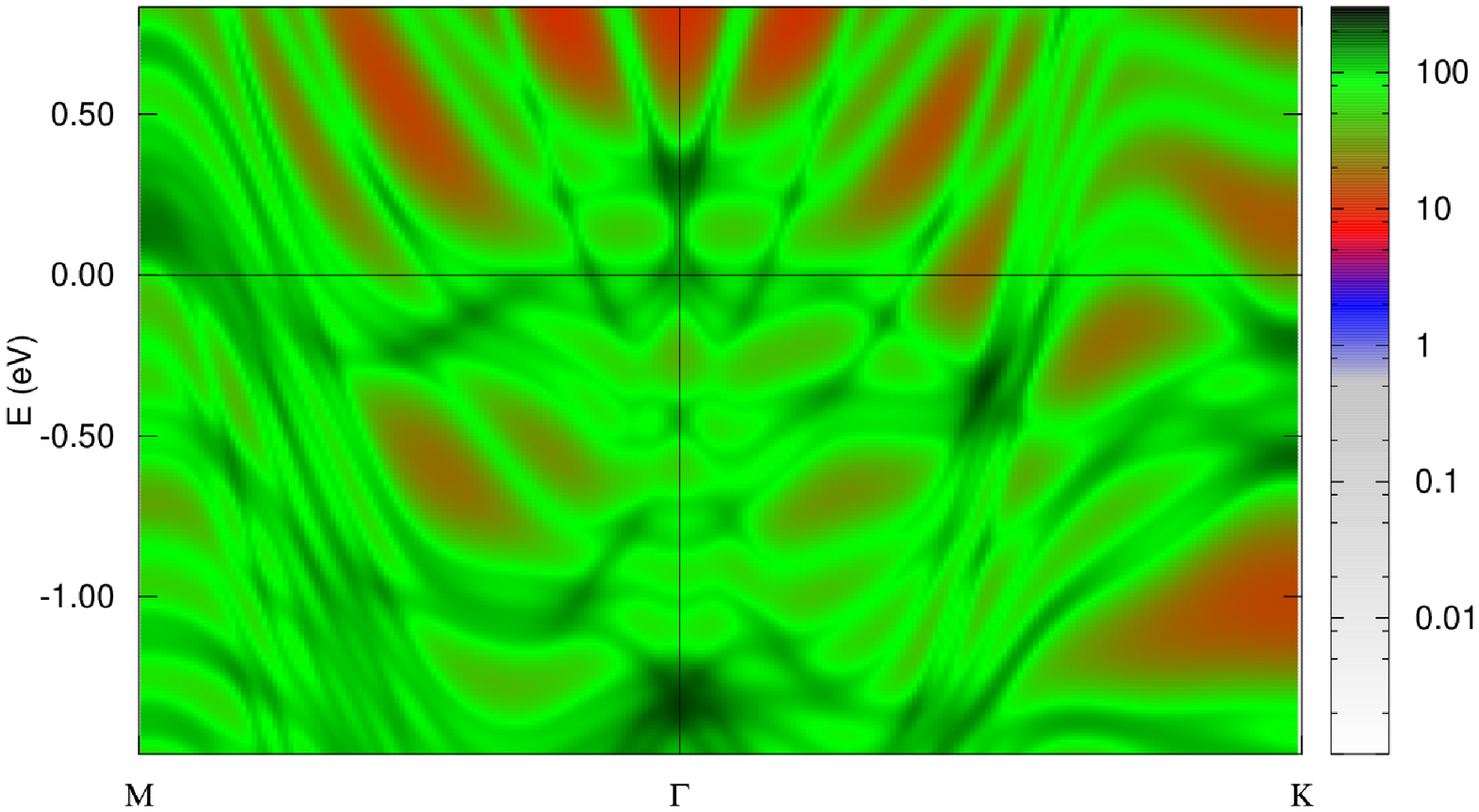}\;(b)
\caption{\label{fig:Fe-Pt_bsf} The Bloch spectral function calculated
  for 1ML Fe/Pt(111) using two values of the imaginary part of the
  energy: 0.1 meV (a) and 5 meV (b). 
 }
\end{figure}

\begin{figure}
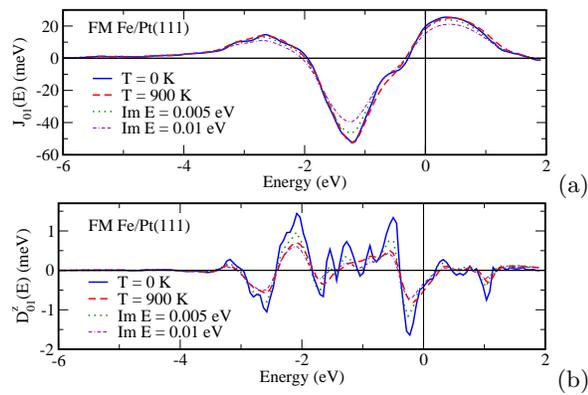

\includegraphics[width=0.4\textwidth,angle=0,clip]{Pt-Fe_J_01_vs_E_and_T_FM.eps}\;(a)
\includegraphics[width=0.4\textwidth,angle=0,clip]{Pt-Fe_DMI_01_vs_E_and_T_FM.eps}\;(b)
\caption{\label{fig:Fe-Pt_JXC_vs_E_vibra}
 The occupation dependent Fe-Fe exchange coupling parameter $J_{01}$ (a)
 and z-component of the DMI $D^{z}_{01}$  in 1ML Fe/Pt(111) calculated
 for two values of the {\em lattice temperature} $T_{\rm lat}$. Dotted
 and dashed-dotted lines  represent the results obtained with the
 imaginary energy part of 5 and  10 meV, respectively.}
\end{figure}

\begin{figure}
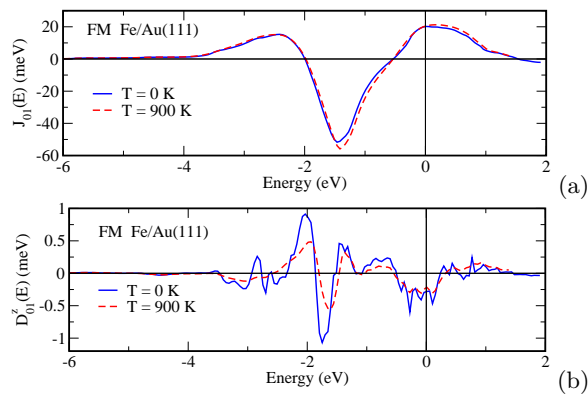

\includegraphics[width=0.4\textwidth,angle=0,clip]{Au-Fe_J_01_vs_E_and_T_FM.eps}\;(a)
\includegraphics[width=0.4\textwidth,angle=0,clip]{Au-Fe_DMI_01_vs_E_and_T_FM.eps}\;(b)
\caption{\label{fig:Au-Pt_JXC_vs_E_vibra}
 The occupation dependent Fe-Fe exchange coupling parameter $J_{01}$ (a)
 and z-component of the DMI $D^{z}_{01}$  in 1ML Fe/Au(111) calculated
 for two values of the {\em lattice temperature} $T_{\rm lat}$.   }
\end{figure}

Finally, as an example for two-dimensional systems, we present results
for 1ML Fe on a Pt (111) and Au (111) substrate, respectively. The  
lack of inversion symmetry leads to non-vanishing Dzyaloshinskii-Moriya
interactions (DMI) in these systems. Therefore we will discuss here the
impact of lattice vibrations not only on the isotropic exchange but also
on the anisotropic  interactions. As the Curie
temperatures evaluated within the MFA are  $\approx 800$~K for
Fe/Pt(111) and $\approx 900$~K for Fe/Au(111), the highest {\em  lattice temperature}
used in our calculations is 900~K.
Figs.\ \ref{fig:FePt_J_DMIx} and \ref{fig:FeAu_J_DMIx} show results for
the Fe-Fe isotropic exchange interaction (a), 
the  $x$- (b) and the $z$-component (c) of the DMI, calculated for the
FM reference state of these systems. As one can see, in both cases a
similar behavior has been found for isotropic exchange interactions $J_{ij}$  
 as a function of the Fe-Fe distance $R_{ij}$ with a weak dependency on the
 lattice temperature.
%
%
On the other hand, the dependence of the DMI components, $D^{\alpha}_{ij}$, 
on thermal lattice vibrations is much more pronounced. 
Interestingly, an opposite trend of the temperature induced
modifications of the $D^{\alpha}_{ij}$ parameters shows up for different
Fe-Fe distances. A similar behavior can also be seen 
when comparing the first-neighbor DMI for the
systems under consideration. While in the case
of 1ML Fe/Au(111) an increasing amplitude of
thermal lattice vibrations results in a decrease of
the Fe-Fe DMI (see Fig.\ \ref{fig:FeAu_J_DMIx} (b), the DMI increases  
with increasing {\em lattice temperature} in the case of  1ML Fe/Pt(111).
To get more insight concerning the influence of lattice vibrations 
on the exchange interactions,
the nearest-neighbor exchange parameters
have been calculated as a function of occupation
for two different  values of the lattice temperature. 
 Figs. \ref{fig:Fe-Pt_JXC_vs_E_vibra} and \ref{fig:Au-Pt_JXC_vs_E_vibra}
 show the isotropic Fe-Fe exchange coupling parameter $J_{01}$ (a) and
 $z$-component of the DMI, $D^{z}_{01}$ (b). For the parameter $J_{01}$ only a weak change 
caused by an increase of the {\em lattice temperature} can be seen over 
whole regime of occupation numbers represented in the figures.
In contrast to this, Figs. \ref{fig:Fe-Pt_JXC_vs_E_vibra} (b)
and \ref{fig:Au-Pt_JXC_vs_E_vibra} show
a very pronounced impact of the lattice vibrations on the 
parameter $D^z_{01}$.
As one can see in the figures, $D^z_{01}$ seen as a function of the
occupation, has a non-monotonous behavior at low temperature with the
observed 'fine structure' associated with avoided crossings of the
energy bands. These details of the electronic
structure can be seen in Fig. \ref{fig:Fe-Pt_bsf} (a)
for 1ML Fe/Pt(111) in the vicinity of the Fermi energy. 
The rapid changes of the DMI occur when
the apparently varied  Fermi level passes through an avoided crossing of
the energy bands (see discussion in \cite{KNA15,San17}).  
The prominent features in the DMI plots seen in
Figs. \ref{fig:Fe-Pt_JXC_vs_E_vibra} (b) and
\ref{fig:Au-Pt_JXC_vs_E_vibra} (b) are created by those energy 
bands that give a dominant contribution to  $D^z_{01}$. 
When the {\em lattice temperature} increases to $T_{\rm lat} = 900$ K, 
the 'fine structure' of $D^z_{01}(E)$ seen as a function of $E$ is
washed out for both systems. Partially, this can be  attributed to a
smearing of the energy bands due to an increasing electron scattering 
by the thermal lattice vibrations. 
This mechanism is demonstrated in Fig. \ref{fig:Fe-Pt_bsf} (b) 
that represents the Bloch spectral function calculated 
for an imaginary part of the energy of 5 meV 
mimicking a decrease of the life time of the electronic
 states connected with the electron scattering by lattice vibrations. 
This modification of the electronic structure leads  
for 1ML Fe/Pt(111) to the changes of  $J_{01}(E)$ and $D^z_{01}(E)$ 
as function of the energy shown in Fig. \ref{fig:Fe-Pt_JXC_vs_E_vibra}
(a) and (b) by dotted lines. 
Dashed-dotted lines represent corresponding 
results obtained for an  imaginary 
part of the energy of 10 meV. 
In the case of the DMI, one can see a
decrease of the amplitude of modulations with energy 
when the imaginary part of the energy increases. However, comparing
these results with the results obtained for $T_{lat} = 900 K$, 
it is obvious that the influence of 
thermal lattice vibrations on the exchange parameters 
also stems to a large extent from their 
impact on the matrix elements given in Eq. (\ref{eq:Jij_ME}).
%

\section{Summary}

To summarize, the alloy analogy model was used to calculate the exchange
coupling parameters taking into account randomly distributed atomic
displacements in the lattice giving access this way 
to temperature induced modifications 
of the exchange parameters. Focusing both
on the isotropic exchange and Dzyaloshinskii-Moriya
 interactions, it is demonstrated that 
-- depending on the material -- the
effect of lattice vibrations on the exchange parameters can be rather
significant and should be taken into account in simulations of
finite-temperature magnetic properties of these systems. Moreover, the
present approach allows to make a corrections to the exchange coupling
parameters in random alloys with alloy components 
having different atomic radius resulting in turn in randomly distributed
atomic displacements, e.g. in high-entropy alloys characterized by
rather significant static 
mean-square atomic displacements \cite{MSW+19}.

 \section{Acknowledgment}
Financial support by the DFG via SFB 1277 (Emergent Relativistic Effects
in Condensed Matter - From Fundamental Aspects to Electronic
Functionality) is gratefully acknowledged.


%

\end{document}